# Optimizing Semi-Stream CACHEJOIN for Near-Real-Time Data Warehousing

**M. Asif Naeem**
*Auckland University of Technology,
Auckland, New Zealand
National University FAST, Islamabad, Pakistan*
mnaeem@aut.ac.nz

**Erum Mehmood**
*University of Management and
Technology,
Lahore, Pakistan*
erum9964@hotmail.com

**M. G. Abbas Malik**
*Universal College of Learning,
Palmerston North, New Zealand*
a.malik@ucol.ac.nz

**Noreen Jamil**
*National University FAST,
Islamabad, Pakistan*
noreen.jamil@nu.edu.pk

**ABSTRACT**

*Streaming data join is a critical process in the field of near-real-time data warehousing. For this purpose, an adaptive semi-stream join algorithm called CACHEJOIN (Cache Join) focusing non-uniform stream data is provided in the literature. However, this algorithm cannot exploit the memory and CPU resources optimally and consequently it leaves its service rate suboptimal due to sequential execution of both of its phases, called stream-probing (SP) phase and disk-probing (DP) phase. By integrating the advantages of CACHEJOIN, in this paper we present two modifications in it. First is called P-CACHEJOIN (Parallel Cache Join) that enables the parallel processing of two phases in CACHEJOIN. This increases number of joined stream records and therefore improves throughput considerably. Second is called OP-CACHEJOIN (Optimized Parallel Cache Join) that implements a parallel loading of stored data into memory while the DP phase is executing. We present the performance analysis of both of our approaches with existing CACHEJOIN empirically using synthetic skewed dataset.*

**INTRODUCTION**

In today's world, the real-time data availability for well-timed and well-informed decisions has become decisive for successful businesses, while data sizes are growing exponentially. Significance of real-time business data devalues, as it gets older. At the same time, the traditional working hours for global enterprises are not germane as they continue to serve customers around the globe and around the clock every day (Golfarelli & Rizzi, 2009), (Vassiliadis, 2009) and (Thomsen & Pedersen, 2005). For uninterrupted global services, continuous real-time data availability for in time business decisions and actions is crucial and indispensable. Traditional offline data-refresh at data warehouses (DWHs) via ETL (Extract-Transform-Load) processes in batch windows (Kimball & Caserta, 2011) are not endurable in this scenario. Therefore, near-real-time data warehousing (NRT-DWH) is an evolving research area and plays a prominent role in supporting cutting-edge and contemporary business strategies and social requirements of the modern era. The modern warehousing techniques are transforming traditional warehouse from a static data repository into an active business entity. This helps to fulfill the contemporary business needs ranging from informing the different stakeholders about latest updates to effective, timely and accurate business decisions.

According to the demand of DWH industry, there is a need to develop an efficient algorithm that performs join operation for bursty and fast streaming data. In NRT-DWH, relational data generated by different data sources needs to reflect in the DWH with a minimal possible delay. Because data is coming from numerous data sources within the organization, it requires significant cleansing and transformation before loading it into the DWH using SQL. Thus, the powerful SQL features can be used to gain consistency and ACID compatibility for join query from the relational schema (Irshad, Yan, & Ma, 2019). ETL processes are used for this purpose (Kimball & Caserta, 2011), (Bornea, Deligiannakis, Kotidis, & Vassalos, 2011). Transformation of extracted data (user sales data) from numerous sources is a crucial phase in ETL processes. In this phase, a stream of new extracted data is joined with a stored data before loading this into the DWH, as shown in Figure 1. Typically, a foreign key from the stream data is joined with the primary key in the master data (Naeem, Dobbie, & Weber, 2012a), (Mokbel, Lu, & Aref, 2004) and (Dittrich, Seeger, Taylor, & Widmayer, 2002). Since the join is between the stream data and the stored data therefore, it is called a semi-stream join.

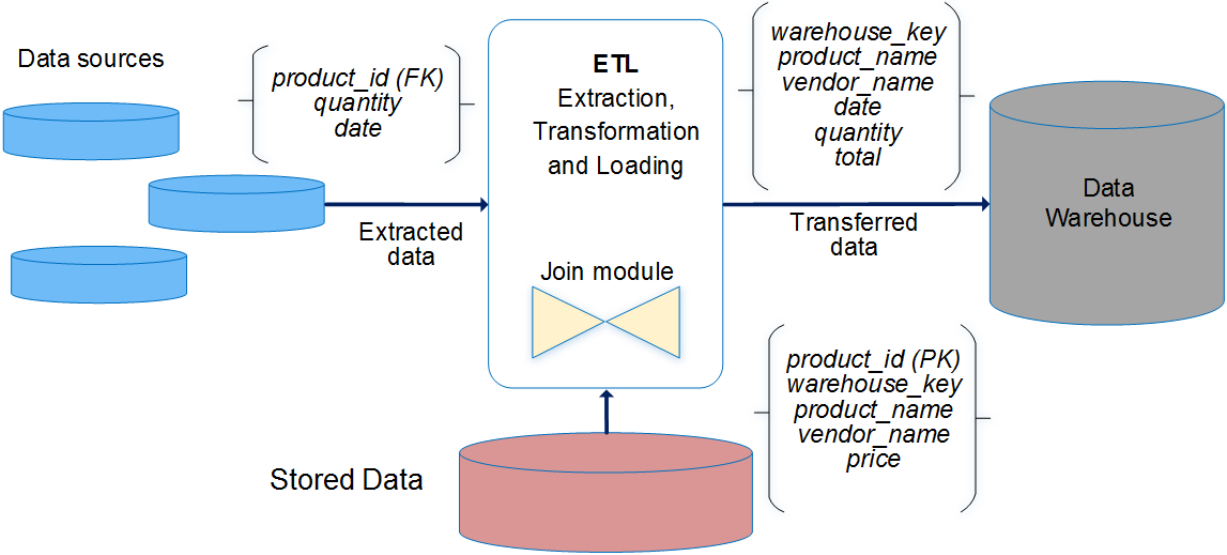

*Figure 1 Illustration of the join during the transformation phase of ETL*

The problem of joining a streaming data with a stored data was first introduced in (Neoklis Polyzotis, Skiadopoulos, Vassiliadis, Simitsis, & Frantzell, 2008) and as a solution a seminal algorithm called MESHJOIN (Mesh Join) was presented. Later, various optimizations in MESHJOIN have been proposed (Bornea et al., 2011), (Naeem et al., 2012a), (Naeem, Dobbie, Weber, & Alam, 2010), (Naeem, Weber, Dobbie, & Lutteroth, 2013) , (Du & Zou, 2013), (Naeem,

Dobbie, & Weber, 2012b). Since the concept of long tail is very common in sales data (Kleinberg, 2002), CACHEJOIN (Naeem et al., 2012a), one of these algorithms, was particularly designed for irregular streams by caching the frequent records of stored data. However, it executes its two SP and DP phases sequentially. Because of the sequential execution, the stream records are waiting unnecessarily before being processed. Thus, the algorithm cannot achieve optimal performance. Parallel execution of the SP and DP phases of CACHEJOIN can significantly speed up the joining process. Further details about limitations of CACHEJOIN are presented later in the paper.

In this paper we propose two modifications in the CACHEJOIN algorithm. First is called P-CACHEJOIN (Parallel Cache Join) (Mehmood & Naeem, 2017)[1] that deals the problem of sequential execution of two phases of CACHEJOIN algorithm. This proposed approach reduces the unnecessary waiting time for the stream records. Second is called OP-CACHEJOIN (Optimized Parallel Cache Join) that introduces an efficient strategy for loading the stored data into memory. This minimizes the disk I/O cost and ultimately improves the service rate. We also want to observe gain in the service rate through each modification separately. Therefore, we evaluated each modification separately.

**RELATED WORK**

Performance optimization of the semi-stream join process for user-update in traditional or near-real-time fashion has been of prime importance for the DWH and database research community. A number of approaches have been published in the literature to optimize the performance of semi-stream join operation. In this section, we present the most relevant from these approaches along with their limitations.

The MESHJOIN algorithm (Neoklis Polyzotis et al., 2008), (N. Polyzotis, Skiadopoulos, Vassiliadis, Simitsis, & Frantzell, 2007) was the first to join a fast stream $S$ coming from various data sources (user-updates) with a large stored data $R$ using limited memory. The algorithm uses two buffers called stream-buffer and disk-buffer to handle inputs coming from the two different sources, *i.e.*, stream and disk. It stores stream records coming from $S$ in a hash table $H$ and their join key values (pointers to the stream records) in a queue $Q$. The primary objective of $Q$ is to keep a record of their arrival order and to make sure the completeness of join for every stream record loaded in memory. For the join process, the algorithm scans $R$ sequentially and cyclically in an infinite loop. Every stream record from $S$ is compared with every record from $R$. Therefore, each stream record remained in the memory for one complete scan-cycle of $R$ hence the size of $R$ remains inversely proportional to the performance of this algorithm. Also, there is no guarantee that even a single stream record would be processed at every scan step of $R$. At the same time, there is a reliance between the number of iterations required to bring $R$ into memory and size of partitions in $Q$ for incoming streams which results into a suboptimal distribution of memory among the join components. Particularly, the size of disk-buffer varies with the size of $R$, which is intuitively incorrect. Additionally, MESHJOIN cannot deal with a non-uniform stream and bursty stream effectively.

---

[1] Preliminary results were presented in the 2nd IEEE International Conference on Cloud computing and Big Data Analysis.

R-MESHJOIN algorithm (Naeem et al., 2010) works similar to MESHJOIN, except it removes the undesired complex dependencies among the components of MESHJOIN algorithm by introducing one additional parameter in the disk-buffer that can vary independently. For example, a size change of $R$ does not affect the size of disk-buffer. The performance of join process improves slightly in R-MESHJOIN, but the problem of dealing non-uniform stream data efficiently still exists in R-MESHJOIN.

MESHJOIN and R-MESHJOIN successfully join $S$ with $R$, but there are a number of factors that require further exploration. First, due to slow sequential access of $R$, disk I/O cost is high as a result the average time of each record in $Q$ is long. Second, these algorithms cannot deal with bursty and non-uniform streams efficiently. Partitioned Join (Chakraborty & Singh, 2009), an optimized version of MESHJOIN, uses a wait buffer for un-matched streams minimizing disk and processing overhead. However, this algorithm observes starvation for infinite stay of un-matched stream records in wait buffer.

The traditional Index Nested Loop Join (INLJ) (Ramakrishnan & Gehrke, 2000) is another choice to join $S$ with $R$. The algorithm uses cluster-based index on the join attribute in $R$. Though the algorithm can handle a bursty stream however, it inputs $S$ record-by-record *i.e.* processes one record against one disk load that decreases the service rate of join significantly.

To overcome the above mentioned limitations Hybrid Join (HYBRIDJOIN) (Naeem, Dobbie, & Weber, 2011), a combination of MESHJOIN and INLJ, was proposed. Unlike to MESHJOIN, the algorithm loads only the useful part of $R$ in the memory using an index. HYBRIDJOIN reduced the stay time for every $S$ record in the join window as well as minimized the disk I/O cost by guaranteeing that at least one $S$ record is processed for each read from $R$. It can also deal with bursty nature of $S$. The major drawback of HYBRIDJOIN is that it cannot handle the skewed distribution of data, like Zipfian distribution (Knuth, 1998) (Anderson, 2006)that commonly occur in the real-world scenarios (Chris, 2006). For example, the study of consumer markets showed that a few products are bought with higher frequency. Thus, records related to these products are frequent in $S$. This problem is addressed in Extended HYBRIDJOIN (X-HYBRIDJOIN) algorithm (Naeem, Dobbie, & Weber) by storing the most frequent part of $R$ in the memory permanently. Contrast to HYBRIDJOIN, X-HYBRIDJOIN divided the disk-buffer in two parts: the first to store frequently used records of $R$, called non-swappable part and the second to load rest of $R$ in partitions based on the index of the oldest record in $Q$, called swappable part. The algorithm significantly reduced the disk-access cost. However, the algorithm executes these two parts of the disk-buffer sequentially that generates some unnecessary wait for the stream records in memory. Moreover, the algorithm stores all the stream records in memory, whether they joined with the swappable or the non-swappable part of the disk-buffer, increasing the cost in terms of loading and unloading the stream records in memory.

Semi-Streaming Index Join (SSIJ) (Bornea et al., 2011) was another recent attempt to join *S* with *R*. SSIJ maximizes the service rate of join process by buffering stream records and dynamically adjusting the available memory space between streams and cached memory blocks.

To improve the performance of join operator another approach was presented that uses a compact data structure for query processing in memory (Vallejos, Caniupan, & Gutierrez, 2018). Another approach has been presented with efficient data structure to process streaming data but the focus was similarity join rather equijoin operation (Wei, Yu, & Lu, 2017).

Di et al. presented a new approach for processing and loading large volumes of data into DWH (Di Tria, Lefons, & Tangorra, 2015). However, the focus of this approach is loading the data in batches and not in near-real-time fashion. Some other studies presented various methodologies for parallel processing of data using data compression techniques (Bellatreche, Cuzzocrea, & Benkrid, 2012), (Pears & Houliston, 2007). The types of data processing include join and aggregate operations. However, the focus is to execute these join and aggregation operations on DWH presentation layer not at ETL layer. Authors in (Hu & Dessloch, 2015) proposed temporal operator models which are responsible for NoSQL temporal data processing. This is not directly related to our approach as the focus is on NoSQL data.

Other literature (Zhao & Siau, 2007), (Triantafillakis, Kanellis, & Martakos, 2004), (Maté et al., 2015), (Trujillo, Luján-Mora, & Song, 2004) and (Choi & Wong, 2009) has been presented to address the issue of data freshness level into DWH. These approaches presented optimized solutions for data refresh rate to improve business intelligence. The studies target both structured as well as unstructured data. Another study (Chee, Yeoh, Gao, & Richards, 2014) provided a framework to facilitate the traceability and accountability of business intelligence products. Candea et al. presented their approach for predicting performance and high query concurrency for data analytics (Candea, Polyzotis, & Vingralek, 2011).

Recently, a semi-stream join algorithm called CACHEJOIN has been proposed to process frequently occurring stream records using SP phase and DP phase which work sequentially. Detailed working of CACHEJOIN is presented in the following section. CACHEJOIN is a much improved and efficient algorithm as compared to all previous studies in terms of non-uniform and skewed data streams. However, two phases of CACHEJOIN do not operate parallel to each other and single disk buffer is used to load disk data partition into memory. We overcome these gaps in this study by developing two modifications in the existing CACHJOIN algorithm. First is called P-CACHEJOIN which executes the SP phase parallel to the DP phase. Second is called OP-CACHEJOIN which loads disk data into two disk buffers parallel to the execution of the DP phase reducing disk I/O cost and improving service rate consequently.

**EXISTING CACHEJOIN AND PROBLEM DEFINITION**

CACHEJOIN (Naeem et al., 2012a) is an adaptive algorithm that was particularly designed to deal with non-uniform stream data. Two hash tables are the fundamental elements of CACHEJOIN with respect to the memory size. One stores *S* records, denoted by $H_S$ while the other stores frequently accessed records of $R$, denoted by $H_R$. The other components of CACHEJOIN are a disk-buffer $D_b$, a queue $Q$ and a stream-buffer $S_b$.

Working of CACHEJOIN algorithm is divided into two sequential phases: SP phase and DP phase. The moment $S$ arrives, SP phase starts its working by looking the stream records into $H_R$, $H_R$ works as intelligent cache mechanism also proposed by (Huang, Lin, & Deng, 2005) to improve the query efficiency for a DWH system in mobile environment. Matched stream records are sent to the output while unmatched records are loaded into $H_S$ by keeping their key values into $Q$. DP phase starts its working when the $H_S$ is completely filled or $S_b$ is empty. In DP phase, a partition of $R$ is loaded into $D_b$ using the oldest record value of $Q$ (from the rear end) as in index. This is the step where CACHEJOIN needs index on $R$. In order to reduce the costly disk access, a number of records from $R$ have to be loaded into $D_b$. The algorithm then probes one-by-one all the records from $D_b$ into $H_S$. In case of match the stream record is generated as an output and at the same time it is also deleted from $H_S$ and $Q$. Because of one-to-many join there can be more than one matches in $H_S$ against one record of $D_b$ therefore, frequency of matched records is counted in DP phase. To decide whether this record of $D_b$ is frequent, the algorithm compares its frequency with a preset threshold value. If the frequency is greater than the threshold value then that record is considered a frequent record and therefore, is moved to $H_R$. Once all records from $D_b$ are probed into $H_S$, DP phase stops its working and the algorithm switches back to SP phase. This one-time sequential execution of both of the phases of CACHEJOIN is called one outer loop iteration of the algorithm.

Since in every iteration, SP phase needs to wait until DP phase finishes its execution and vice versa, the execution of the both phases of the algorithm are inefficient. Thus, resulting the algorithm as a suboptimal solution. Our investigation shows that if both phases of the algorithm run in parallel, the waiting time of stream records can be minimized. Our results, presented later in the paper, show that the service rate significantly improves if parallel execution of these two phases is implemented. Furthermore, if the retrieval of $R$ is implemented parallel to the join operation in DP phase, the major cost (i.e., disk I/O cost) of the algorithm can be reduced drastically.

To address the identified gap, we propose two modifications in the existing CACHEJOIN algorithm. First is called P-CHACHEJOIN (Parallel Cache Join) that overcomes the sequential execution problem of CACHEJOIN. Second is called OP-CACHEJOIN (Optimized Parallel Cache Join) that improves the retrieval of $R$. Our experimental results show that the service rate is improved significantly in case of the both algorithms. The next two sections present the proposed P-CACHEJOIN and OP-CACHEJOIN with their cost models and experimental analysis.

## P- CACHEJOIN

In this modification, we enable the both phases (SP and DP) of CACHEJOIN to execute in parallel. For that we introduce an intermediate-buffer that contains the stream records that are not matched in SP phase. This allows the parallel execution of the both phases which is not the case in existing CACHEJOIN. The detailed description of P-CACHEJOIN is presented in the coming section.

## Architecture

Figure 2 presents the memory architecture and execution layout of P-CACHEJOIN algorithm. Similar to CACHEJOIN major components of P-CACHEJOIN are two hash tables, $H_S$ (stores stream records) and $H_R$ (stores records from $R$), stream-buffer $S_B$, disk-buffer $D_B$ and queue $Q$. In order to run DP and SP phases in parallel, a new component called intermediate-buffer $I_B$ is introduced in P-CACHEJOIN. $I_B$ stores the stream records that do not find their match in $H_R$. Due to this DP phase works in parallel to SP phase by fetching stream records from $I_B$. The role of $I_B$ is as a buffer for DP phase rather a caching module. Furthermore, no join operation is performed on the stream records while these are remained in $I_B$.

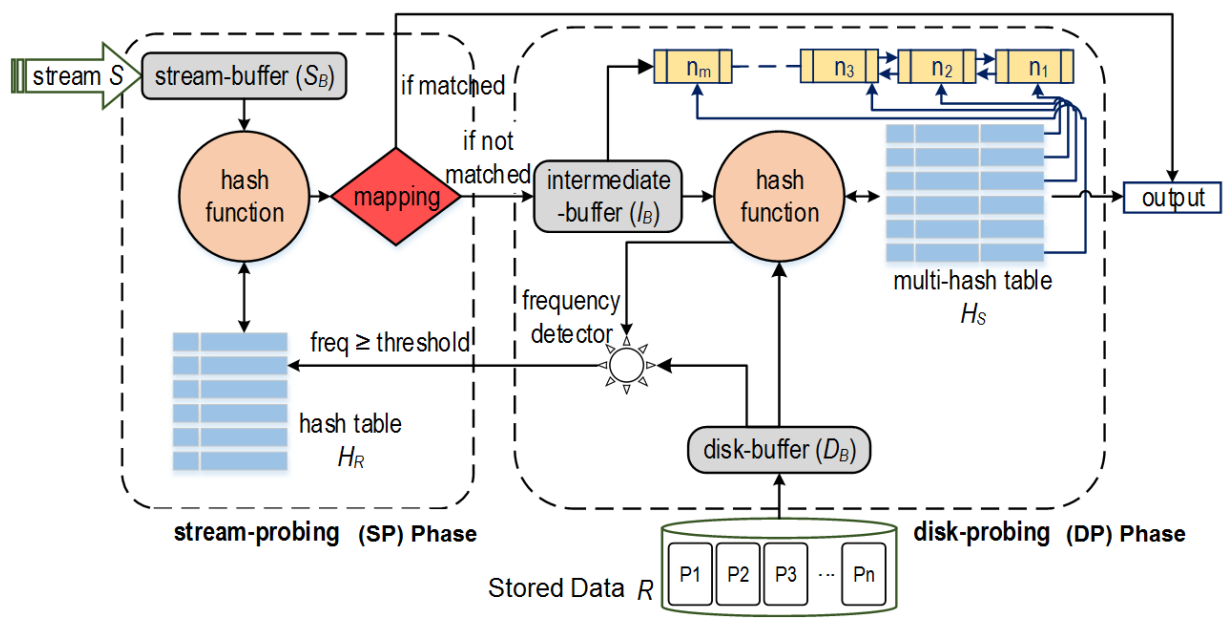

Figure 2 *Architecture of P-CACHEJOIN algorithm*

Unlike CACHEJOIN in P-CACHEJOIN the two hash-join phases run in parallel. In SP phase, the algorithm reads stream records from $S_B$ and looks them up in $H_R$ for their matches. If a match is found, join process is applied and the output is generated. While unmatched stream records are stored in $I_B$. In parallel, DP phase of P-CACHEJOIN keeps fetching the unmatched stream records from $I_B$ and storing them in $H_S$ with their join attribute values in $Q$ for further matching. $I_B$ is always accessible for both of these parallel running phases and keeps updating continuously. After DP phase has fetched the stream records from $I_B$, its remaining execution including frequency detection procedure is same as in CACHEJOIN. By including $I_B$ between two parallel running phases of P-CACHEJOIN, the algorithm performs join operation without an unnecessary pause, which consequently increases the service rate. SP phase stops only if $I_B$ is full while DP phase stops only if $I_B$ is empty. However, these conditions occur rarely.

## Cost Model for P-CACHEJOIN

In this section, we develop the cost model for our P-CACHEJOIN algorithm. The cost model presented here follows the style used in CACHEJOIN. Most of the P-CACHEJOIN algorithm cost model notations are similar to CACHEJOIN with a few additional notations. The notations used in the cost model are given in

The memory for each component in P-CACHEJOIN can be calculated as follows:

Memory(bytes) for $D_b = d_B \cdot v_R$
Memory(bytes) for $H_R = h_R \cdot v_R$
Memory(bytes) for $I_b = i_B \cdot v_S$
Memory(bytes) for $H_S = \alpha(M - (d_B + h_R) \cdot v_R - i_B \cdot v_S)$
where M is the total used memory.
Memory for Q (bytes) $= (1 - \alpha)(M - (d_B + h_R) \cdot v_R - i_B \cdot v_S)$

By accumulating all calculated memories above, the overall memory $M$ used by P-CACHEJOIN can be computed using Eq. (1), given below:

$$M = (d_B + h_R) \cdot v_R + i_B \cdot v_S + \alpha(M - (d_B + h_R) \cdot v_R - i_B \cdot v_S) + (1 - \alpha)(M - (d_B + h_R) \cdot v_R - i_B \cdot v_S) \qquad \text{Eq. (1)}$$

Memory for $S_B$ is not included in the memory cost because it is very small and negligible. For our experiments, 0.05 MB memory is used for $S_B$.

**Processing cost.** To calculate the processing cost of P-CACHEJOIN, we first calculate the individual processing cost for each component as follows:

*Time to read $d_B$ records into $D_B$ (nano secs)* $= c_{I/O}(d_B)$
*Time to look up $\omega_N$ records in $H_R$ (nano secs)* $= \omega_N \cdot c_H$
*Time to look up $d_B$ records in $H_S$ (nano secs)* $= d_B \cdot c_H$
*Time to detect the frequncy of all records in $D_B$ (nano secs)* $= d_B \cdot c_F$
*Time to generate the output for $\omega_N$ records (nano secs)* $= \omega_N \cdot c_O$
*Time to generate the output for $\omega_S$ records (nano secs)* $= \omega_S \cdot c_O$
*Time to read $\omega_N$ records from $S_B$ (nano secs)* $= \omega_N \cdot c_S$
*Time to read $\omega_S$ records from $I_B$ (nano secs)* $= \omega_S \cdot c_S$
*Time to append $\omega_S$ records in $H_S$ and Q (nano secs)* $= \omega_S \cdot c_A$
*Time to delete $\omega_S$ records from $H_S$ and Q (nano secs)* $= \omega_S \cdot c_E$
*Time to append $\omega_S$ records in $I_B$ (nano secs)* $= \omega_S \cdot c_{Ab}$

By aggregating the above execution times, the total processing time of P-CACHEJOIN for one loop iteration can be calculated using Eq. (2). Since the stream records while waiting in I_B are not using any CPU, we do not include this in the processing costs. Also, in our case, since DP phase runs in parallel to SP phase, the waiting time for each record in I_B is very minimal and therefore, we do not evaluate this in our experimentations.

$$C_{loop}(secs) = 10^{-9}[C_{I/O}(d_B) + d_B \cdot (c_H + c_F) + \omega_S \cdot (c_O + c_S + c_A + c_E + c_{Ab}) + \omega_N \cdot (c_H + c_O + c_S)] \qquad \text{Eq. (2)}$$

Since algorithm takes $c_{loop}$ seconds to process $\omega_N$ and $\omega_S$ records from $S$, the service rate $\mu$ can be calculated using Eq. (3).

$$\mu = \frac{(\omega_N + \omega_S)}{c_{loop}} \quad \text{Eq. (3)}$$

Table 1. The cost of execution of P-CACHEJOIN is modeled in terms of memory cost, representing the total memory used by the algorithm, and processing cost, total time of execution for one loop iteration of the algorithm. We describe the both as below.

**Memory cost.** As described above a major portion of total memory is assigned to the two hash tables, $H_S$ and $H_R$. On the other hand, comparatively a much smaller amount of memory is assigned to $Q$, $D_B$, and $I_B$.

The memory for each component in P-CACHEJOIN can be calculated as follows:

Memory(bytes) for $D_b = d_B \cdot v_R$
Memory(bytes) for $H_R = h_R \cdot v_R$
Memory(bytes) for $I_b = i_B \cdot v_S$
Memory(bytes) for $H_S = \alpha(M - (d_B + h_R) \cdot v_R - i_B \cdot v_S)$
where $M$ is the total used memory.
Memory for $Q$ (bytes) $= (1 - \alpha)(M - (d_B + h_R) \cdot v_R - i_B \cdot v_S)$

By accumulating all calculated memories above, the overall memory $M$ used by P-CACHEJOIN can be computed using Eq. (1), given below:

$$M = (d_B + h_R) \cdot v_R + i_B \cdot v_S + \alpha(M - (d_B + h_R) \cdot v_R - i_B \cdot v_S) + (1 - \alpha)(M - (d_B + h_R) \cdot v_R - i_B \cdot v_S) \quad \text{Eq. (1)}$$

Memory for $S_B$ is not included in the memory cost because it is very small and negligible. For our experiments, 0.05 MB memory is used for $S_B$.

**Processing cost.** To calculate the processing cost of P-CACHEJOIN, we first calculate the individual processing cost for each component as follows:

Time to read $d_B$ records into $D_B$ (nano secs) $= c_{I/O}(d_B)$
Time to look up $\omega_N$ records in $H_R$ (nano secs) $= \omega_N \cdot c_H$
Time to look up $d_B$ records in $H_S$ (nano secs) $= d_B \cdot c_H$
Time to detect the frequncy of all records in $D_B$ (nano secs) $= d_B \cdot c_F$
Time to generate the output for $\omega_N$ records (nano secs) $= \omega_N \cdot c_O$
Time to generate the output for $\omega_S$ records (nano secs) $= \omega_S \cdot c_O$
Time to read $\omega_N$ records from $S_B$ (nano secs) $= \omega_N \cdot c_S$
Time to read $\omega_S$ records from $I_B$ (nano secs) $= \omega_S \cdot c_S$
Time to append $\omega_S$ records in $H_S$ and $Q$ (nano secs) $= \omega_S \cdot c_A$
Time to delete $\omega_S$ records from $H_S$ and $Q$ (nano secs) $= \omega_S \cdot c_E$
Time to append $\omega_S$ records in $I_B$ (nano secs) $= \omega_S \cdot c_{Ab}$

By aggregating the above execution times, the total processing time of P-CACHEJOIN for one loop iteration can be calculated using Eq. (2). Since the stream records while waiting in I_B are not using any CPU, we do not include this in the processing costs. Also, in our case, since DP phase runs in parallel to SP phase, the waiting time for each record in I_B is very minimal and therefore, we do not evaluate this in our experimentations.

$$C_{loop}(secs) = 10^{-9}[C_{I/O}(d_B) + d_B \cdot (c_H + c_F) + \omega_S \cdot (c_O + c_S + c_A + c_E + c_{Ab}) + \omega_N \cdot (c_H + c_O + c_S)] \quad \text{Eq. (2)}$$

Since algorithm takes $c_{loop}$ seconds to process $\omega_N$ and $\omega_S$ records from S, the service rate $\mu$ can be calculated using Eq. (3).

$$\mu = \frac{(\omega_N + \omega_S)}{c_{loop}} \quad \text{Eq. (3)}$$

*Table 1: Notations used in cost models of P-CACHEJOIN and OP-CACHEJOIN*

| Parameter Name | Notation |
|---|---|
| No. of stream records processed in each iteration through $H_R$ | $\omega_N$ |
| No. of stream records processed in each iteration through $H_S$ | $\omega_S$ |
| Stream record size (*bytes*) | $v_S$ |
| Disk record size (*bytes*) | $v_R$ |
| Size of $D_B$ (*records*) | $d_B$ |
| Size of $H_R$ (*records*) | $h_R$ |
| Size of $H_S$ (*records*) | $h_S$ |
| Size of $I_B$ (*records*) | $i_B$ |
| Memory load for $H_S$ | $\alpha$ |
| Memory load for $Q$ | $1 - \alpha$ |
| Time to read $d_B$ disk records into $D_b$ (*nano secs*) | $c_{I/O}(d_B)$ |
| Time to look-up one record in $H_S$ (*nano secs*) | $c_H$ |
| Time to generate the output for one record (*nano secs*) | $c_O$ |
| Time to remove one record from $H_S$ and $Q$ (*nano secs*) | $c_E$ |
| Time to read one stream record from $S_b$ or $I_B$ (*nano secs*) | $c_S$ |
| Time to append one record in $H_S$ and $Q$ (*nano secs*) | $c_A$ |
| Time to detect the frequency of one disk record in $D_b$ (*nano secs*) | $c_F$ |
| Time to append one record in $I_B$ (*nano secs*) | $c_{Ab}$ |
| Total time for one iteration (secs) | $c_{loop}$ |

## Tuning

The cost model for P-CACHEJOIN can be used to adjust $i_B$ to obtain the optimal performance. To find the optimal $i_B$ we ran the different experiments by varying $i_B$ using total memory of 100 MB and $d_B$ of 850 records. Service rate $\mu$ was measured for each memory setting of $i_B$. Figure 3 demonstrates the impact of different values of $i_B$ on the service rate. By increasing $i_B$ greater than 2 MB, service rate of P-CACHEJOIN starts decreasing. It means, if we include more unmatched records in $I_B$ it decreases the share of memory for $H_S$ and $Q$. Considering the above observation, we assign 2 MB to $I_B$ is used in our subsequent experiments.

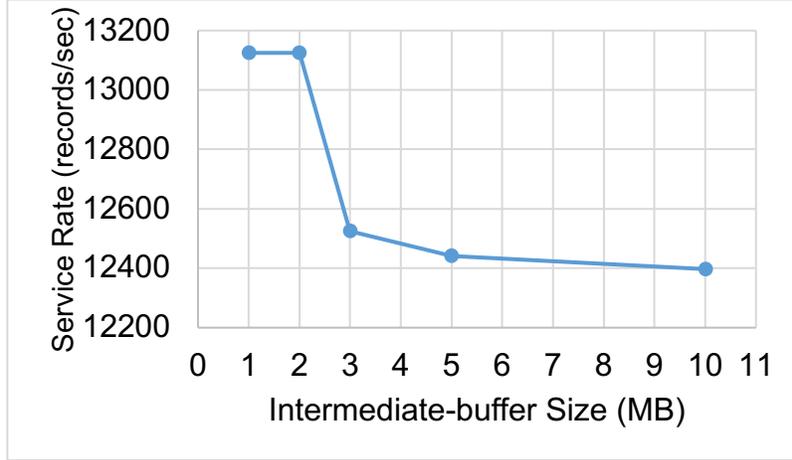

*Figure 3: Performance analysis of P-CACHEJOIN for different values of $i_B$*

## Experimental Settings

We implemented P-CACHEJOIN algorithm in Java using Eclipse IDE with the following specifications:

**Hardware and Software Specifications.** We executed our experiments on a multi-core processor, *i.e.*, Core-i5 with a frequency of 1.70 GHz for each core. $R$ is stored using MySQL database. Fetch size for the ResultSet is set equal to $d_B$, *i.e.*, 850 records. Apache plugins and Java API (nanoTime) are used to record different measurements that are required to calculate the costs of P-CACHEJOIN.

**Data Specifications.** The performance of P-CACHEJOIN has been analyzed using synthetic data. The stream data is generated at run time using stream generating script of CACHEJOIN [8], based on Zipfian law [20]. In our experiments, the size of $R$ is varied from 0.5 million records to 2 million records to evaluate the effect of size of $R$ on the performance of P-CACHEJOIN algorithm. The size of each record in $R$ is 120 bytes while the size of each record in the stream data is 20 bytes. The size of $I_B$ is 2 MB. A join attribute value stored in the queue $Q$ is of 4 bytes and the value of the fudge factor for the multi-hash-map $H_S$ is 8.

**System of measurement.** The performance of the P-CACHEJOIN is measured using service rate $\mu$, that represents the number of records processed in one second. The measurements are recorded regularly after some iterations of the loop. For each setting, we measured readings for sufficient numbers of iterations (minimum 1000 iterations) and their average is used in the final calculations. Moreover, it is assumed that no other applications are running in parallel during the execution of the algorithm.

## Performance Evaluation

To test the behavior of the algorithm, we used three different parameters. These three parameters are: the size of $R$, the total available memory $M$, and the exponent of the Zipfian distribution. For the sake of brevity, the discussion over each parameter's effect has been restricted to the one-dimensional variation, *i.e.*, only one parameter has been varied at a time.

**Performance comparisons for varying size of R.** In this experiment, we varied the size of *R* and measured the performance of both P-CACHEJOIN and CACHEJOIN algorithms. While the values of the other parameters are fixed, *i.e.*, Zipfian exponent is set to 1 and the total available memory *M* is set to 50 MB. We used the discrete sizes of *R* in order to evaluate the effects of size of *R* on the performance. (a) shows the results of our experiments and it clearly demonstrates that for all different sizes of *R*, P-CACHEJOIN performs significantly better than CACHEJOIN. The factor of improvement is more visible for small sizes of *R* e.g. in case of 50MB.

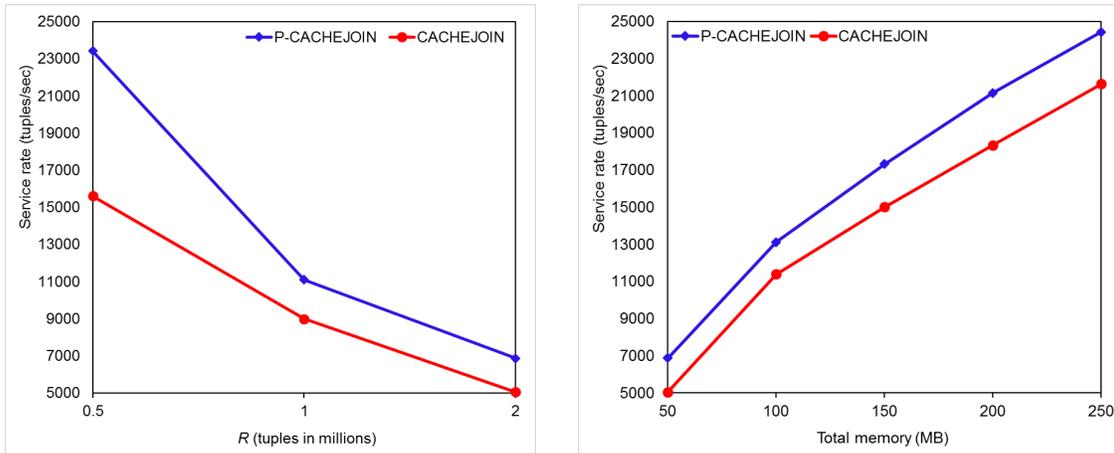

(a) *Varying size of R*        (b) *Varying memory budget*

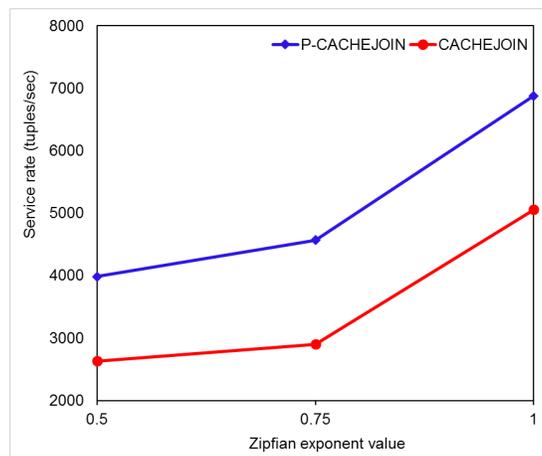

(c) *Varying Zipfian exponent*
Figure 4. Service rates analysis

**Performance comparisons for different memory budgets.** To evaluate the effect of different memory budgets on the performances of P-CACHEJOIN and CACHEJOIN, we fixed the values of other two parameters, i.e., size of *R* to 2 million records and Zipfian's exponent to 1, while changing the memory budget from 50 MB to 250 MB. Figure 4(b) shows the results and it clearly validate that P-CACHEJOIN performs always better than CACHEJOIN for each memory size.

**Performance comparisons while varying skew in S.** Finally, we evaluated the performance of P-CACHEJOIN and CACHEJOIN by varying the skew in *S*. To vary the skew, the value of the

Zipfian exponent has been varied from 0.5 to 1 with an interval of 0.25. At 0.5, the input stream $S$ has less skew while at 1, it has high skew. The results, given in Figure 4(c), shows that P-CACHEJOIN again performs better than CACHEJOIN for all the values of skew.

## OP-CACHEJOIN

The major cost in P-CACHEJOIN algorithm is the disk I/O cost of loading $R$ into the memory. In P-CACHEJOIN algorithm, DP phase needs to wait until the new data from $R$ is available in $D_B$. This caused an un-necessary delay in processing of the fast-incoming $S$. Thus, by optimally reading $R$ to $D_B$ can improve the service rate of the join algorithm. Based on this observation, we modified our P-CACHEJOIN algorithm by introducing another disk-buffer. This enables the parallel loading of data from $R$ to the new disk-buffer while the algorithm is performing join with the older disk-buffer. The modified version is called OP-CACHEJOIN (Optimized Parallel Cache Join).

In OP-CACHEJOIN the new disk-buffer is denoted by $D_{B1}$ while the old disk-buffer is denoted by the same symbol $D_B$, given in P-CACHEJOIN. The join attributes values at the front and rear of the $Q$ serve as the index values for loading data from $R$ to $D_{B1}$ and $D_B$ respectively. The central idea behind these two disk-buffers is that we can load one disk-buffer with the new records from $R$, while the other disk-buffer is being used by DP phase. Once DP phase finishes its joining process using one buffer, it switches to the other disk-buffer with almost no delay, if the buffer is ready. By using these two disk-buffers, we overcome the limitations of P-CACHEJOIN and further optimize the service rate.

### Architecture

The executions architecture of OP-CACHEJOIN is shown in Figure 5.

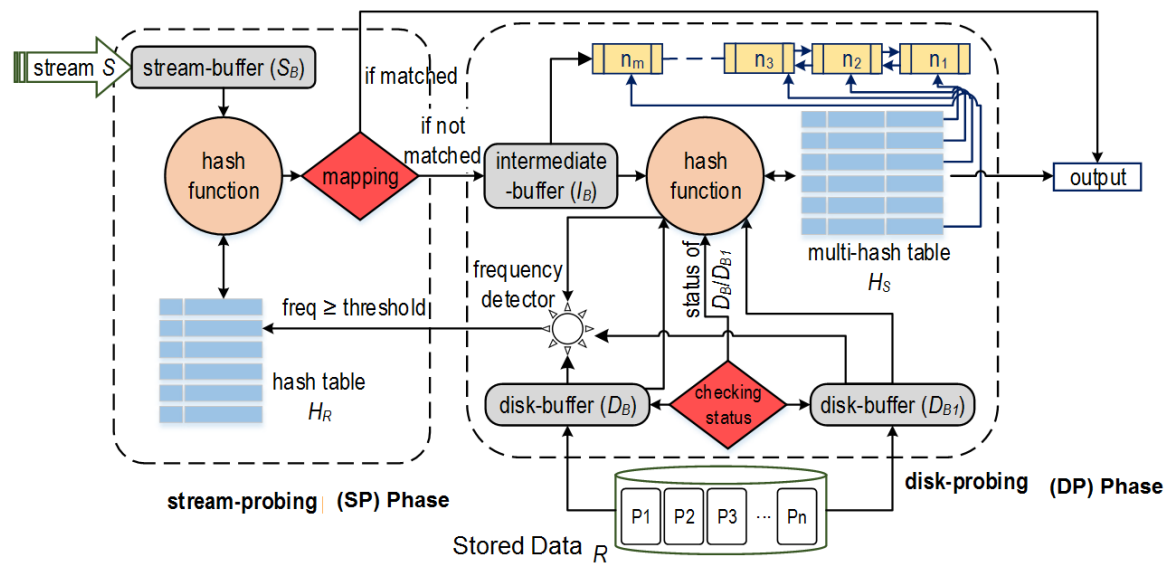

Figure 5. Data structures and architecture of OP-CACHEJOIN

To achieve independent execution for loading the disk-buffers with records from $R$, we set each buffer with status busy, empty, and full so the buffer with empty status can be refilled while the

DP phase is in action. Two separate threads are used to load data to the both disk-buffers. A disk-buffer is ready for DP phase only when its status is 'full'. As soon DP phase finishes its working with $D_B$, its status changes from 'busy' to 'empty' that also activates the relevant thread to refill $D_B$ with new data. Meanwhile the algorithm switches to $D_{B1}$. Once, the new data is loaded to $D_B$, its status changes from 'empty' to 'full'. Rest of the working of the OP-CACHEJOIN for its SP and DP phases is same as the P-CACHEJOIN algorithm.

By adding one more disk-buffer and making disk-buffer loading phase parallel to DP phase, we observed a remarkable improvement in the service rate.

## Cost Model for OP-CACHEJOIN

In this section, we develop the cost of the proposed OP-CACHEJOIN algorithm both in terms of memory and processing costs. Since the size of $D_{B1}$ is equal to the size of $D_B$ given in P-CACHEJOIN therefore, we denote the size of each buffer by $d_B$ records. For both memory and processing costs we only calculate the costs for $D_B$ and $D_{B1}$ and based on these we modify Equation (1) and (2).

**Memory cost.**
Memory for $D_B$ and $D_{B1}$ (bytes) $= 2d_B \cdot v_R$

By adding the above cost in Equation (1) the memory cost for OP-CACHEJOIN can be calculated.

**Processing cost.** Similar to memory cost, we first calculate the cost for loading records from $R$ to $D_B$ and $D_{B1}$ as follows:

$$Cost\ to\ read\ d_B\ disk\ tuples\ into\ D_B\ or\ D_{B1}\ (nano\ secs) = \frac{1}{2}c_{I/O}(d_B)$$

An important thing to note in case of OP-CACHEJOIN is that since the algorithm fills both of the disk-buffers in parallel while running DP phase so the disk I/O cost for one loop iteration (i.e. against one DP phase) is almost half of the I/O cost given in P-CACHEJOIN. Therefore, the total cost for one iteration of OP-CACHEJOIN can be calculated only by changing $C_{I/O}(d_B)$ into $\frac{1}{2}c_{I/O}(d_B)$ using Equation (2). This is the reason why OP-CACHEJOIN outperforms P-CACHEJOIN. Also, in each loop iteration of DP phase since the algorithm probes records from the one disk-buffer therefore we count the probing cost against the one disk-buffer.

## Performance Evaluation

Similar to P-CACHEJOIN algorithm, OP-CACHEJOIN algorithm has also been tested by varying the same three parameters: the size of $R$, The total available memory $M$, and the exponent of the Zipfian distribution.

**Performance comparisons for varying size of $R$.** First, we varied the size of $R$ and measured the performance of OP-CACHEJOIN, P-CACHEJOIN, and CACHEJOIN algorithms. Figure 6(a) shows the results of our experiments and it is clearly depicted that OP-CACHEJOIN significantly outperformed the other two.

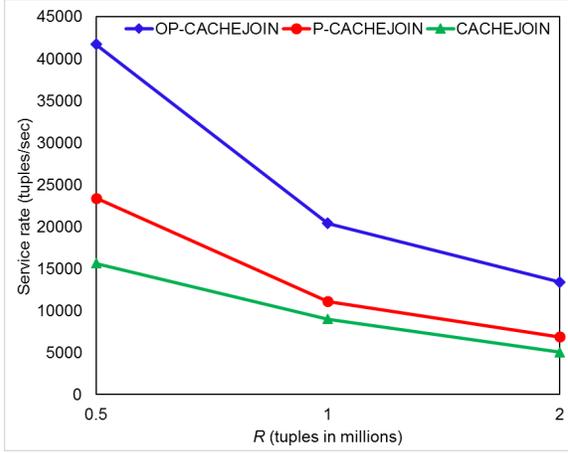
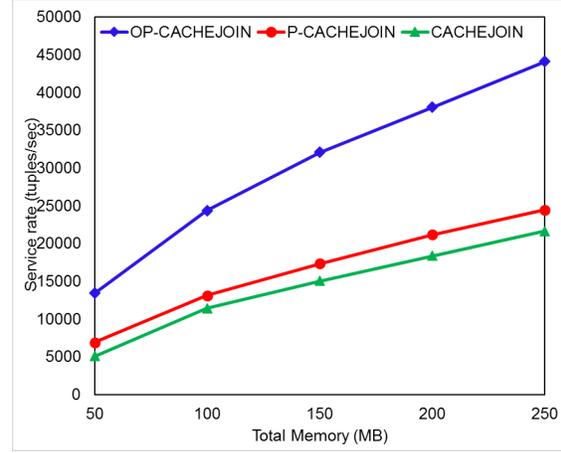

(a) *Varying size of R*

(b) *Varying memory budget*

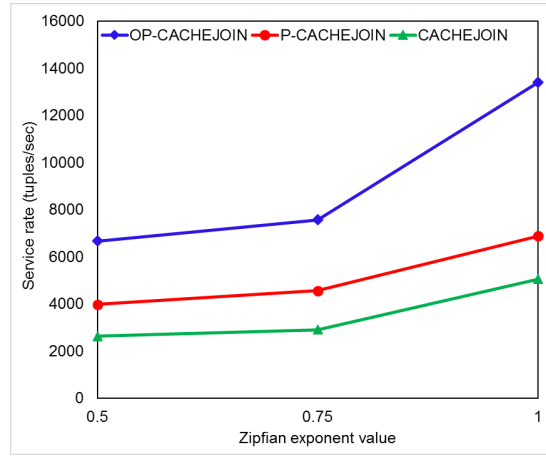

(c) *Varying Zipfian exponent*

Figure 6. Service rates analysis

**Performance comparisons for different memory budgets.** Similar to P-CACHEJOIN evaluation, in this experiment we evaluated the effect of different memory sizes on the service rate of all the three algorithms. We varied the memory size from 50 MB to 250 MB while fixed the values for other parameters, $R$ is equal to 2 million records and Zipfian's exponent is equal to 1. Figure 6(b) shows the results of our experiment and it is again obvious that OP-CACHEJOIN outperformed both P-CACHEJOIN and CACHEJOIN algorithms by a huge margin.

**Performance comparisons while varying skew in $S$.** Finally, we evaluated the performance of OP-CACHEJOIN by varying the skew in $S$. As mentioned earlier, to vary the skew, the value of the Zipfian exponent has been varied from 0.5 to 1 with an interval of 0.25. Figure 6(c) demonstrates the results of our experiment and again the service rate has been improved significantly in case of OP-CACHEJOIN which validates our arguments.

**THEORETICAL AND PRACTICAL IMPLICATIONS**

The primary purpose of business intelligence is to improve the quality of making decisions with minimal latency in data processing. The proposed OP-CACHEJOIN is a semi-stream join algorithm that achieves efficient service rate as compare to the state-of-the-art approaches in this area. The new algorithm has two-fold advantages over the existing CACHEJOIN. First, the both phases of the algorithm run in parallel. Second, the algorithm uses efficient approach to load disk-based master data into memory using two buffers. This incurs less execution time to process streaming data which helps in making timely business decisions.

Our approach is effective for all those organizations which maintain their DWHs on near-real-time-basis and need fresh data to make their business decisions. Example of these organisations can be stock-exchange business, weather prediction systems, and online personalised marketing.

**CONCLUSIONS AND FUTURE DIRECTIONS**

In this paper, we explored a recent algorithm called CACHEJOIN (Cache Join). CACHEJOIN is an adaptive algorithm for irregular stream data, however, the algorithm cannot exploit the memory and CPU resources optimally due to the sequential execution of both of its SP and DP phases. Consequently, it does not achieve an optimal service rate. To overcome these short comings, we presented two modifications in the existing CACHEJOIN algorithm. First is called P-CACHEJOIN (Parallel Cache Join) that allows parallel execution of the both phases of CACHEJOIN. Therefore, unlike the existing CACHEJOIN algorithm, for each iteration SP phase does not wait for the ending of DP phase and vice versa. This optimized the utilization of available memory and CPU recourse and thus resulted in improving the service rate significantly. Although, P-CACHEJOIN algorithm improves the performance for the join process, but still the major cost has been the loading of the stored data into the disk-buffer due to its slow access rate. To solve this problem, we proposed our second modification, called OP-CACHEJOIN (Optimized Parallel Cache Join). In OP-CACHEJOIN, we introduced a new disk-buffer that allows the parallel loading of data from the master data to each disk-buffer while the algorithm is executing its DP phase. It scans the master data almost twice in size with the same value of I/O cost in P-CACHEJOIN. Our experiments have shown a significant improvement of OP-CACHEJOIN over the P- CACHEJOIN and CACHEJOIN algorithms. We also developed the cost models for the both of our modifications. In future, we have a plan to deploy the parallel execution of the two phases on separate physical computers. This will further accelerate the performance of join operation.

**REFERENCES**

Bellatreche, L., Cuzzocrea, A., & Benkrid, S. (2012). Effectively and efficiently designing and querying parallel relational data warehouses on heterogeneous database clusters: The f&a approach. *Journal of Database Management (JDM), 23*(4), 17-51.

Bornea, M. A., Deligiannakis, A., Kotidis, Y., & Vassalos, V. (2011). *Semi-Streamed Index Join for near-real time execution of ETL transformations*.

Candea, G., Polyzotis, N., & Vingralek, R. (2011). Predictable performance and high query concurrency for data analytics. *The VLDB Journal—The International Journal on Very Large Data Bases, 20*(2), 227-248.
Chakraborty, A., & Singh, A. (2009). *A partition-based approach to support streaming updates over persistent data in an active datawarehouse*.
Chee, C.-H., Yeoh, W., Gao, S., & Richards, G. (2014). Improving business intelligence traceability and accountability: An integrated framework of BI product and metacontent map. *Journal of Database Management (JDM), 25*(3), 28-47.
Choi, R. H., & Wong, R. K. (2009). Efficient Filtering of Branch Queries for High-Performance XML Data Services. *Journal of Database Management (JDM), 20*(2), 58-83.
Chris, A. (2006). The long tail: Why the future of business is selling less of more. *Hyperion*.
Di Tria, F., Lefons, E., & Tangorra, F. (2015). Benchmark for approximate query answering systems. *Journal of Database Management (JDM), 26*(1), 1-29.
Dittrich, J.-P., Seeger, B., Taylor, D. S., & Widmayer, P. (2002). *Progressive merge join: A generic and non-blocking sort-based join algorithm*.
Du, W., & Zou, X. (2013). *The algorithm of the join data stream with diskresident relation*.
Golfarelli, M., & Rizzi, S. (2009). A survey on temporal data warehousing. *International Journal of Data Warehousing, 5*.
Hu, Y., & Dessloch, S. (2015). Temporal data management and processing with column oriented nosql databases. *Journal of Database Management (JDM), 26*(3), 41-70.
Huang, S.-M., Lin, B., & Deng, Q.-S. (2005). Intelligent cache management for mobile data warehouse systems. *Journal of Database Management (JDM), 16*(2), 46-65.
Irshad, L., Yan, L., & Ma, Z. (2019). Schema-Based JSON Data Stores in Relational Databases. *Journal of Database Management (JDM), 30*(3), 38-70.
Kimball, R., & Caserta, J. (2011). *The Data Warehouse? ETL Toolkit: Practical Techniques for Extracting, Cleaning, Conforming, and Delivering Data*: John Wiley & Sons.
Kleinberg, J. (2002). *Bursty and hierarchical structure in streams*.
Knuth, D. E. (1998). *The art of computer programming: sorting and searching* (Vol. 3): Pearson Education.
Maté, A., Llorens, H., de Gregorio, E., Tardío, R., Gil, D., Munoz-Terol, R., & Trujillo, J. (2015). A novel multidimensional approach to integrate big data in business intelligence. *Journal of Database Management (JDM), 26*(2), 14-31.
Mehmood, E., & Naeem, M. A. (2017). *Optimization of cache-based semi-stream joins.* Paper presented at the 2017 IEEE 2nd International Conference on Cloud Computing and Big Data Analysis (ICCCBDA).
Mokbel, M. F., Lu, M., & Aref, W. G. (2004). *Hash-Merge Join: A Non-blocking Join Algorithm for Producing Fast and Early Join Results*, Washington, DC, USA.
Naeem, M. A., Dobbie, G., & Weber, G. (2011). *X-HYBRIDJOIN for near-real-time data warehousing*, Berlin, Heidelberg.
Naeem, M. A., Dobbie, G., & Weber, G. (2011). HYBRIDJOIN for Near-Real-Time Data Warehousing. *International Journal of Data Warehousing and Mining, 7*(4), 21-42.
Naeem, M. A., Dobbie, G., & Weber, G. (2012a). *A lightweight stream-based join with limited resource consumption*.
Naeem, M. A., Dobbie, G., & Weber, G. (2012b). *Optimised X-HYBRIDJOIN for near-real-time data warehousing*.

Naeem, M. A., Dobbie, G., Weber, G., & Alam, S. (2010). *R-MESHJOIN for Near-real-time Data Warehousing*, Toronto, Canada.

Naeem, M. A., Weber, G., Dobbie, G., & Lutteroth, C. (2013). *SSCJ: A semi-stream cache join using a front-stage cache module*.

Pears, R., & Houliston, B. (2007). Optimization of multidimensional aggregates in data warehouses. *Journal of Database Management (JDM), 18*(1), 69-93.

Polyzotis, N., Skiadopoulos, S., Vassiliadis, P., Simitsis, A., & Frantzell, N. (2008). Meshing Streaming Updates with Persistent Data in an Active Data Warehouse. *IEEE Trans. on Knowl. and Data Eng., 20*(7), 976-991. doi:http://dx.doi.org/10.1109/TKDE.2008.27

Polyzotis, N., Skiadopoulos, S., Vassiliadis, P., Simitsis, A., & Frantzell, N. E. (2007). *Supporting Streaming Updates in an Active Data Warehouse*, Istanbul, Turkey.

Ramakrishnan, R., & Gehrke, J. (2000). *Database management systems*: McGraw-Hill.

Thomsen, C., & Pedersen, T. B. (2005). *A survey of open source tools for business intelligence*.

Triantafillakis, A., Kanellis, P., & Martakos, D. (2004). Data warehousing interoperability for the extended enterprise. *Journal of Database Management (JDM), 15*(3), 73-84.

Trujillo, J., Luján-Mora, S., & Song, I.-Y. (2004). Applying UML and XML for designing and interchanging information for data warehouses and OLAP applications. *Journal of Database Management (JDM), 15*(1), 41-72.

Vallejos, C., Caniupan, M., & Gutierrez, G. (2018). Compact Data Structures to Represent and Query Data Warehouses into Main Memory. *IEEE Latin America Transactions, 16*(9), 2328-2335.

Vassiliadis, P. (2009). A survey of Extract–transform–Load technology. *International Journal of Data Warehousing and Mining (IJDWM), 5*(3), 1-27.

Wei, H., Yu, J. X., & Lu, C. (2017). String similarity search: A hash-based approach. *IEEE Transactions on Knowledge and Data Engineering, 30*(1), 170-184.

Zhao, L., & Siau, K. (2007). Information mediation using metamodels: An approach using XML and common warehouse metamodel. *Journal of Database Management (JDM), 18*(3), 69-82.

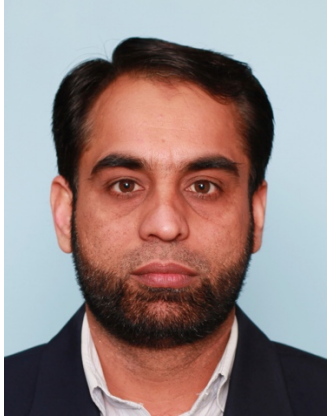

**Muhammad Asif Naeem**

Dr Muhammad Asif Naeem is Director of Data Science Research Group and Senior Lecturer in School of Engineering, Computer and Mathematical Sciences, Auckland University of Technology, Auckland, New Zealand. He received his PhD degree in Computer Science from The University of Auckland, New Zealand. He has been awarded with a best PhD thesis award from The University of Auckland. Before that Asif has done his Master's degree in Computer Science with distinction. He has about fifteen years research, industrial and teaching experience. He has published over 50 research papers in high repute journals, conferences, and workshops in his area. His recent research has been published in Information Systems and in Journal of Computational and Applied Mathematics which are ranked A* and A respectively in Computing Research and Education Association (CORE). He has been reviewing for well-known journals and conferences in his area. He is organising an IEEE workshop IWDS since 2013. He is an IEEE member. His research interests are Data Stream Processing, Real-time Data Warehousing, Big Data Management, Knowledge Engineering, and Data Science.

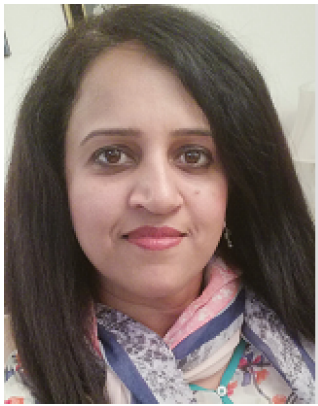

**Erum Mehmood**

Erum Mehmood is a PhD student at University of Management and Technology, Lahore Pakistan. She received her MPhil degree in computer science from NCBA&E Lahore Pakistan, in 2017. Her MPhil dissertation is in the area of stream processing for real-time data warehousing. She is currently working as lecturer in computer science department at Government Degree College Lahore, Pakistan. Her research interests include big data analytics, stream processing, ETL, and real-time data warehousing.

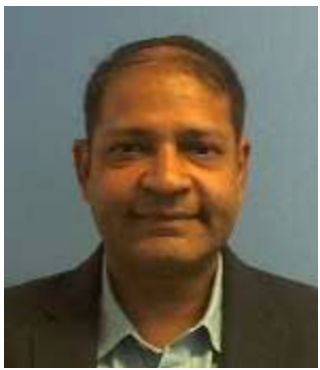

**M G Abbas Malik**

Dr M G Abbas Malik has done his Ph.D. in Computer Science from University of Grenoble, France. He has also done his Master in Computational Linguistics and Master in Computer Science from University of Paris 7 – Denis Didrot, France and University of the Punjab, Pakistan respectively. His research interest included Artificial Intelligence, Machine Learning, Data Mining, Data Analytics and Natural Language Processing. Software development, web and mobile app development are one of his strong points. He has been serving as Assistant Professor and Senior Lecturer in various academic institutes of France, Pakistan, Saudi Arabia and New Zealand. He is currently working as senior academic member at School of Business and ICT in Universal College of Learning, Palmerston North, New Zealand.

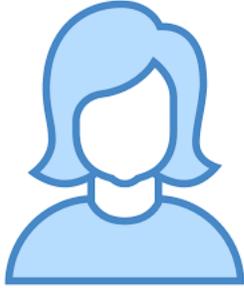
**Noreen Jamil**
Dr Noreen Jamil received her PhD degree in Computer Science from The University of Auckland, New Zealand. As a part of her PhD programme, she was a visiting research fellow at Department of Mathematics, University of Maryland, USA. She has more than 10 years academic and research experience at university level. She has published more than 18 research papers in well-reputed conferences and journals including in IEEE and ELSEVIER. Recently she has published 2 papers in Journal of Computational and Applied Mathematics, one of the world leading Computational Mathematics Journals. She has received the best paper award in IEEE-ICDIM 2013 and the best student paper award at the University of Auckland in 2013. Her research interests include Computational Mathematics, Human Computer Interaction, Numerical Computation, and Constraint Programming.